\relax
\documentclass{article}
\usepackage{ijcai21}

\usepackage{times}
\usepackage{soul}
\usepackage{url}
\usepackage[hidelinks]{hyperref}
\usepackage[utf8]{inputenc}
\usepackage[small]{caption}
\usepackage{graphicx}
\usepackage{helvet}
\usepackage{amsthm}
\usepackage{booktabs}
\usepackage{algorithm}
\usepackage{algorithmic}
\usepackage{subcaption}
\usepackage{amsmath, amssymb}
\usepackage{subcaption}
\usepackage[dvipsnames]{xcolor}
\usepackage{latexsym}
\newcommand{\ra}[1]{\renewcommand{\arraystretch}{#1}}
\newtheorem{theorem}{Theorem}
\newtheorem{proposition}{Proposition}

\newtheorem{definition}{Definition}
\def\tuple#1{( #1 )}

\title{Manipulation of k-Coalitional Games on Social Networks\thanks{This work has been supported in part by the Israel Science
Foundation under grant 1958/20, the Ministry of Science, Technology \& Space, Israel and the EU project TAILOR under
Grant 992215.} }
\author{

    Naftali Waxman,
    Noam Hazon,
     Sarit Kraus
}

\begin{document}
\maketitle
\begin{abstract}
In many coalition formation games the utility of the agents depends on a social network. In such scenarios there might be a manipulative agent that would like to manipulate his connections in the social network in order to increase his utility. We study a model of coalition formation in which a central organizer, who needs to form $k$ coalitions, obtains information about the social network from the agents.
The central organizer has her own objective: she might want to maximize the utilitarian social welfare, maximize the egalitarian social welfare, or simply guarantee that every agent will have at least one connection within her coalition. 
In this paper we study the susceptibility to manipulation of these objectives, given the abilities and information that the manipulator has. Specifically, we show that if the manipulator has very limited information, namely he is only familiar with his immediate neighbours in the network, then a manipulation is almost always impossible. Moreover, if the manipulator is only able to add connections to the social network, then a manipulation is still impossible for some objectives, even if the manipulator has full information on the structure of the network. On the other hand,  if the manipulator is able to hide some of his connections, then all objectives are susceptible to manipulation, even if the manipulator has limited information, i.e., when he is familiar with his immediate neighbours and with their neighbours.

\end{abstract}

\section{Introduction}
Coalition formation is one of the fundamental research problems in multi-agent systems \cite{chalkiadakis2011computational}. 
Broadly speaking, coalition formation is concerned with partitioning a population of agents into disjointed teams (or coalitions) with the aim that some system-wide performance measure is maximized. Indeed, in many coalition formation games there is a central organizer that would like to maximize some objective.

\begin{table*}[t]
\centering
\ra{1}
\begin{tabular}{@{}lcccc@{}}\toprule& \multicolumn{2}{c}{$Directed$} &  \multicolumn{2}{c}{$Undirected$}\\
\cmidrule{2-3} \cmidrule{4-5}
 & $Add$ & $Remove$ & $Add$ & $Remove$\\\midrule
\textbf{Max-Util} &
Strict(F~\ref{fig:Util_add}) & Strict(F~\ref{fig:Util_dir_remove}) & Strict(F~\ref{fig:Util_add})& Strict(F~\ref{fig:Util_remove_Improve})*\\
\textbf{Max-Egal} &
Strategyproof(T~\ref{thrm:egal_dir_add})& Strict(F~\ref{fig:Egal_directed_remove}) & LB,UB(F~\ref{fig:Egal_undirected_add_LB},\ref{fig:Egal_undirected_add_UB}), W-proof(T~\ref{thrm:egal_undir_add}) & Strict(F~\ref{fig:Egal_undirected_remove})\\
\textbf{At-least-1} &
Strategyproof(T~\ref{thrm:least1_add_dir}) & LB(F~\ref{fig:least1_remove}), UB-Proof(T~\ref{thrm:least1_remove}) & Strict(F~\ref{fig:Least1_undirected_add})& LB(F~\ref{fig:least1_remove}),UB-proof(T~\ref{thrm:least1_remove})\\
\bottomrule
\end{tabular}
\caption{Summary of the results.
The parentheses near a result refer to the corresponding figure (F) or theorem (T). The results hold for both full information and distance 2, except for the result with the * , which holds only for the full information case. Key: LB/UB/Strict = the objective is subject to LB/UB/Strict-improvement, W-proof = the objective is weak-proof.
}
\label{tbl:summary}
\end{table*}
One assumption that is usually made is that the utility function of the agents is known and given as an input. However, in some real-world scenarios the organizer obtains the information regarding the utility function directly from the agents. For example, when dividing students into classes, it is a common practice to ask the students about their social relationships~\cite{alonhighschool}, since a student is more satisfied if the number of friends she has within the class to which she is assigned is maximized. Similarly, when assigning workers to tasks, a manager would be interested in the interpersonal relationships between potential team members.
Ideally, the agents would report their true social relationships so that the organizer will be able to choose the most appropriate
coalition structure. 
However, there might be scenarios in which an agent is better off manipulating the organizer by misreporting his relationships.


Indeed, the problem of manipulation in the context of coalitional games has been studied recently ~\cite{wright2015mechanism,flammini2017strategyproof}. These studies have looked for strategyproof mechanisms for forming the coalitions, at the cost of non-optimal social welfare (SW).
In this paper we propose a complementary approach. We study in which situations there might be an agent with an incentive to manipulate the organizer, and in which situations no agent has an incentive to manipulate the organizer, and thus a special strategyproof mechanism is not needed (see \cite{vallee2014study} for a similar approach). 
This analysis is in the same vein as the works of \cite{gibbard1973manipulation} and \cite{satterthwaite1975strategy} in the context of voting, that studied in which situations there might be a voter with an incentive to misreport her true vote.


We focus on $k$-coalitional games, where exactly $k$ coalitions must be formed \cite{sless2018forming}. We assume that the agents' utilities depend on a social network that represents the social relationships among the agents. Specifically, the social network is modeled as an unweighted graph where the vertices are agents and the edges indicate friendship among the agents. The utility function of an agent is the number of friends she has within the coalition to which she is assigned. Actually, our model is a special case of simple Additively Separable Hedonic Games (ASHGs)~\cite{bogomolnaia2002stability}.
In addition, there is an organizer that would like to maximize some objective function, thus she needs to obtain the structure of the social network from the agents' reports regarding their friendships. 
In such situations, it is possible that one manipulative agent would like to misreport his friendship connections, in order to increase his utility. In particular, a manipulator may hide some of his connections or he may add connections by reporting fake connections (with agents with which the manipulator does not have real connections).

Within these settings we study different objectives for the organizer and analyze their susceptibility or resistance to manipulation. Specifically, we study the objective of maximizing the utilitarian social welfare (Max-Util), maximizing the egalitarian social welfare (Max-Egal), and the At-least-1 objective, where the organizer is only interested in guaranteeing that every agent will have at least one friendship connection within her coalition.
Moreover, we study different settings regarding the abilities and information that the manipulator has. Specifically, we study a manipulator that is able to report fake friendship connections (i.e. add edges) and a manipulator that is able to hide some of his friendship connections (i.e., remove edges). We study the situation in which the manipulator has full information regarding the structure of the network and situations in which the manipulator has limited information: he may be familiar only with his connections to his immediate neighbours in the social network (denoted distance $1$), or he may also be familiar with their connections to other agents (denoted distance $2$). (In both scenarios the manipulator knows about the existence of all agents in the social network, but not how they are connected.)

Table \ref{tbl:summary} summarizes our results for the full information and distance $2$ settings.
Overall, all of the objectives are susceptible to manipulation by removing edges, even in the case of distance $2$. On the other hand, in some settings there are objectives that are resistant to manipulation by adding edges, even in the case of full information.
Note that the results for distance $1$ do not appear in the Table, since in almost all of the cases the objectives are resistant to manipulation.

\begin{figure}
 \includegraphics[page=1]{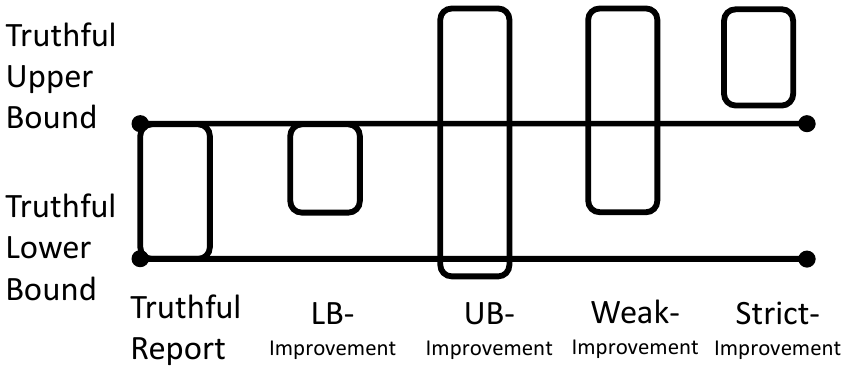}
     \caption{The possible utility values of the manipulator for each manipulation type.}
 \label{fig:manipulation_types}
\end{figure}

\section{Related Work}


There are several studies that developed strategyproof mechanisms for forming coalitions. 
\citeauthor{dimitrov2004enemies} \shortcite{dimitrov2004enemies} discussed ASHGs where agents have both positive and negative edges, and provided a strategyproof algorithm for finding stable outcomes.
\citeauthor{rodriguez2009strategy} \shortcite{rodriguez2009strategy} analyzed strategyproof core stable solutions' properties. They showed that single lapping rules are necessary and sufficient for the existence of a unique core-stable partition.
\citeauthor{aziz2013pareto} \shortcite{aziz2013pareto} showed that the serial dictatorship mechanism is strategyproof with appropriate restrictions over the agents' preferences. 
\citeauthor{flammini2017strategyproof} \shortcite{flammini2017strategyproof} focused on the utilitarian SW in ASHGs and Fractional Hedonic Games, and proposed strategyproof mechanisms at the cost of non-optimal social welfare.
\citeauthor{wright2015mechanism} \shortcite{wright2015mechanism} studied a model of ASHG that is very similar to our model, but instead of restricting the number of coalitions they restricted the size of each coalition. Within their model they proposed a strategyproof mechanism that achieves good and fair experimental performance, despite not having a theoretical guarantee.
All of these works have looked for strategyproof mechanisms, while our approach is to study in which situations a strategyproof mechanism is indeed needed, and in which situations it is not needed since manipulation is impossible.

Our approach is similar to the approach of 
\citeauthor{vallee2014study} \shortcite{vallee2014study}, who studied general hedonic games and Sybil attacks, i.e., manipulations, by adding false agents to the game. \citeauthor{vallee2014study} showed that hedonic games with Nash stability as the solution concept are very robust to Sybil attacks, but when contractual individual stability is the solution concept then every game is manipulable. 
Recently, \citeauthor{alonhighschool} \shortcite{alonhighschool} considered the At-least-1 objective, and analyzed whether a group of manipulators can guarantee being in the same coalition in every game. \citeauthor{alonhighschool} showed that such manipulation is almost always impossible.



\section{Definitions}
\label{sec:defn}
Let $A=\{a_{1},\ldots ,a_{n}\}$ be a finite, non-empty set of agents, and let $G=\tuple{A,E}$ be a graph with no self loops, representing the friendship connections between the agents.
The set of immediate neighbours of $a_{i}$ in $G$ is denoted by $N(a_{i})$.
We also refer to $G$ as the \emph{social network}. A coalition $C\subseteq A$ is a subset of agents; we do not require that agents in a coalition form a connected component in the corresponding social network.
Let $u(a_i,C)$ be the utility that agent $a_i$ would obtain from being in coalition $C$. This value is simply the sum of edges corresponding to the immediate neighbors of $a_i$ that are members of $C$. That is,
$ u(a_{i},C)=|C\cap N(a_{i})|$.

We assume that there is a central organizer that would like to partition the agents into $k$ coalitions in order to satisfy some objective, $obj$.
Let $\Pi_k$ denote the set of partitions of $A$ that contain exactly $k$ non-empty subsets where $0 < k \leq n$. 
We refer to elements of $\Pi_k$ as \emph{coalition structures} (CS), and typically use $P, P', \ldots$ to denote such coalition structures.
We assume that the utility of agent $a$ depends only on the members of her coalition. Therefore, if $P \in \Pi_{k}$, $C\in P$, and $a\in C$, then we use the notation $u(a,P)$ to refer to $u(a,C)$.  
Note that there may be several coalition structures that satisfy a given objective $obj$. We denote this set of coalitions as $O_{obj}(G) \subseteq \Pi_k$ and refer to them as \emph{solutions}. In many cases we  omit the reference to $obj$ when it is clear.

In our setting the social network is formed based on self reports of the agents. That is, each agent $a_i$ is asked by the organizer to list all of her friendship connections. Formally, let $R=\{r_{1},\ldots, r_{n}\}$ be the set of reports, where $r_{i}\subseteq A\setminus\{a_{i}\}$.
In such a scenario there might be a manipulative agent $m$. We begin by assuming that $m$ has full information regarding the social network (we relax this assumption in Section~\ref{sec:limited}) and the objective of the central organizer, $obj$.
Moreover, we assume that $m$ is able to misreport his friendship connections and thus add non-existing edges connecting him to other agents or omit existing edges between him and other agents. 
We denote these two types of manipulators by $m^+$ and $m^-$, respectively. We did not consider a manipulator that is capable of both adding and removing edges since it does not add any new results: in all of our objectives only one capability is needed to show susceptibility to manipulation.
Let $G^m=\tuple{A,E^m}$ be the resulting social network known to the organizer after the manipulation $r_m$, and let $N^m(a)$ be the set of immediate neighbours of $a$ in $G^m$.
Note that if $G$ is directed then we assume that $m$ is able to add or remove only outgoing edges, i.e., add an edge $(m,a_i) \notin E$ or remove an edge $(m,a_i) \in E$. $m$ is not able to add or remove incoming edges, i.e., add an edge $(a_i,m) \notin E$ or remove an edge $(a_i,m) \in E$. That is, $N(a)=N^m(a)$ for every agent $a \neq m$. 
If $G$ is undirected then a manipulation by adding edges is relevant when the organizer adds an edge $(a_i,a_j)$ to $G$ if either $a_i\in r_j$ or $a_j\in r_i$. On the other hand, a manipulation by removing edges is relevant when the organizer adds an edge $(a_i,a_j)$ to $G$ only when both $a_i\in r_j$ and $a_j\in r_i$.
When the context is clear we will sometimes refer to $G^m$ as the manipulation.
 


Clearly, the goal of the manipulator is a successful manipulation. Indeed, in our setting there are several ways to define what a successful manipulation is, since there may be several coalition structures that satisfy $obj$ in $G$ and in $G^m$, but the utility of $m$ might be different in each such coalition structure. Formally, given a network $G$ and objective $obj$:

\begin{definition}
\label{defn:lower_bound_improvement}
A manipulation $r_m$ is a \emph{lower bound improvement} (LB-improvement) for a manipulator $m$ if:
\[
\underset{P\in O_{obj}(G^{m})}{\min}(u(m,P))>\underset{P\in O_{obj}(G)}{\min}(u(m,P)).
\]
%

\label{defn:upper_bound_improvement}
A manipulation $r_m$ is an \emph{upper bound improvement} (UB-improvement) for a manipulator $m$ if:
\[
\underset{P\in O_{obj}(G^{m})}{\max}(u(m,P))>\underset{P\in O_{obj}(G)}{\max}(u(m,P)).
\]
\end{definition}
That is, LB-improvement eliminates coalition structures with low utility for the manipulator, while UB-improvement adds coalition structures with higher utility for the manipulator. For example, assume that for an objective $obj$ and a graph $G$ there are two possible CSs. That is, $O_{obj}(G)=\{P_1,P_2\}$. Moreover, assume that $u(m,P_1)=1$ and $u(m,P_2)=2$. If there exists a manipulation $r_m$ where $O_{obj}(G^m)=\{P_2\}$ (or any other $P$ satisfying $u(m,P)=2$) then $G^m$ is a LB-improvement. If there exists a manipulation $r_m$ where $O_{obj}(G^m)=\{P_1,P_3\}$ and $u(m,P_3)=3$ then $G^m$ is an UB-improvement.
LB-improvement can be considered risk aversion of some sort, while UB-improvement suits an optimistic manipulator looking for higher utilities.

There is a stronger variant of manipulation which is both LB- and UB-improvement. An even stronger variant is where every coalition structure is strictly better than every possible coalition structure that would have been generated with $m$'s true preferences. Formally:
\begin{definition}
\label{improvement}
A manipulation $r_m$ is a \emph{weak-improvement} for a manipulator $m$ if it is both LB-improvement and UB-improvement for him.
\label{defn:strict_improvement}
A manipulation $r_m$ is a \emph{strict-improvement} for a manipulator $m$ if:
\[
\underset{P\in O_{obj}(G^{m})}{\min}(u(m,P))>\underset{P\in O_{obj}(G)}{\max}(u(m,P)).
\]
\end{definition}

Revisiting our example, a manipulation where $O_{obj}(G^m) = \{P_3\}$ is a strict-improvement. Note that the utility $u(m,P)$ is always calculated over the original graph $G$ with the manipulator's true neighbours.
We refer to the different manipulations: LB, UB, weak, and  strict-improvement, as \emph{manipulation types}.
Finally, we define the susceptibility and resistance to a manipulation type of a given objective.
\begin{definition}
\label{defn:subject_to_improvement}
An objective $obj$ is \emph{subject to LB-improvement} by manipulator $m$ over (un)directed networks
if there exists a (un)directed social network $G$ and a manipulation $r_m$ such that $r^m$ is a LB-improvement for $m$.
Otherwise, we say that $obj$ is \emph{LB-proof} against $m$.
\end{definition}
The definitions for the other manipulation types are similar.
When an objective is both LB- and UB-proof, we say that it is \emph{strategyproof}. Figure~\ref{fig:manipulation_types} demonstrates the possible utility values of the manipulator for each manipulation type.

\section{Full Information}
\label{sec:full_info}
\begin{figure}[t]
    \centering  
        \begin{subfigure}{0.09\textwidth}
            \centering
            \includegraphics[page=3,width=\textwidth]{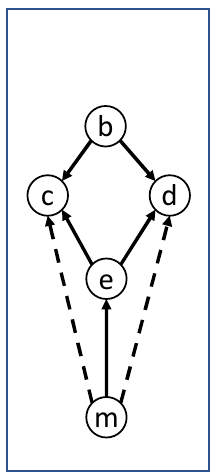}
            \caption{\\U/S}
            \label{fig:Util_add}
        \end{subfigure}
        \hfill
        \begin{subfigure}{0.09\textwidth}
            \centering
            \includegraphics[page=12,width=\textwidth]{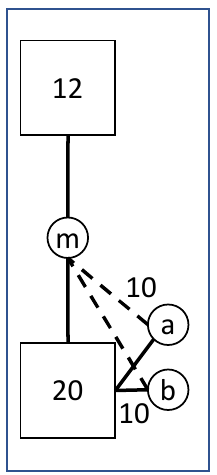}
            \caption{\\E/LB}
            \label{fig:Egal_undirected_add_LB}
        \end{subfigure}
        \hfill
        \begin{subfigure}{0.09\textwidth}
            \centering
            \includegraphics[page=15,width=\textwidth]{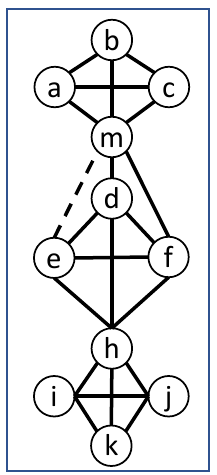}
            \caption{\\E/UB}
            \label{fig:Egal_undirected_add_UB}
        \end{subfigure}
        \hfill
        \begin{subfigure}{0.09\textwidth}
            \centering
            \includegraphics[page=49,width=\textwidth]{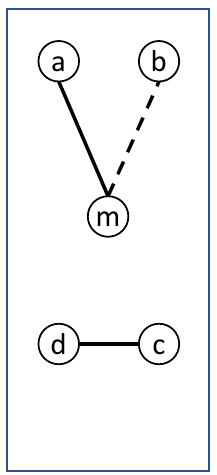}
            \caption{\\1/S}
            \label{fig:Least1_undirected_add}
        \end{subfigure}

    \caption{Figures for the proofs of manipulations by $m^+$. Key:U=Max-Util, E=Max-Egal, 1=At-least-1, S=strict-improvement, UB=UB-improvement, LB=LB-improvement}
    \label{fig:add_graphs}
\end{figure}

We begin our analysis of the objectives and their susceptibility or resistance to the different types of manipulation. 
To show susceptibility to manipulation, we provide figures that depict the scenarios in which manipulation is possible. We use $k=2$ in all of our proofs, but they can easily be extended for any $k$. We use the following notations: In all of the figures the vertex $m$ represents the manipulator. A node is represented by a circle, and a rectangle with a number $X$ represents a clique of $X$ agents. An edge going to (from) a clique represents edges going to (from) all nodes in the clique. An edge going to (from) a clique with a number $X$ represents $X$ edges going to (from) arbitrarily chosen $X$ nodes in the clique. 
If the graph is directed, then an undirected edge $(a,b)$ represents two directed edges, $(a,b)$ and $(b,a)$. If we prove a result regarding an undirected graph and refer to a figure with a directed graph then every directed edge represents an undirected edge. 
Overall, Figure~\ref{fig:add_graphs} provides scenarios for $m^+$ and Figure~\ref{fig:remove_graphs} provides scenarios for $m^-$. Therefore, the dotted edges in Figure~\ref{fig:remove_graphs} are the fake edges that are added by the manipulator, while the dotted edges in Figure~\ref{fig:remove_graphs} are the edges that are removed by the manipulator.

We note that almost all of the susceptibility results in the full information setting (except for Proposition~\ref{Prop:util_full_info}) are derived by the results of susceptibility in the distance 2 setting (see Section~\ref{subsecdist-2}). Therefore, in this section we mostly provide the results regarding resistance to manipulation.

\subsection{Max-Util}
\label{sec:utilirarain}

Maximizing the utilitarian social welfare (Max-Util) is a very common objective in hedonic games \cite{aziz2015welfare}. It was also studied from the perspective of graph theory, since finding a CS (with $k$ coalitions) that maximizes the utilitarian SW is equivalent to finding a minimum $k$-cut \cite{branzei2009coalitional}.  
Utilitarian SW is defined as the sum of the utilities of all agents. Formally, it is $\underset{a\in A}{\sum}u(a,P)$.

Max-Util is always susceptible to manipulation; in all of the situations that we consider, this objective is subject to strict-improvement. 
Recall that our susceptibility results are derived from the distance 2 setting. However, there is one situation in which the susceptibility to manipulation in the distance 2 setting is not known 
, and thus we show that even in this situation Max-Util is subject to strict-improvement. 

\begin{proposition}
\label{Prop:util_full_info}
Max-Util is subject to strict-improvement by a manipulator $m^-$ over an undirected network.
\end{proposition}
\begin{proof}
Consider the network $G$ as depicted in Figure~\ref{fig:Util_remove_Improve}. Recall that $k=2$. Clearly, the minimum $2$-cut is obtained by cutting the upper clique ($\{a,b,c,d,e,f\}$) from the rest of the network, yielding a minimum $2$-cut of size 3. The manipulator's utility is thus 5. By removing the dotted edges, the minimum $2$-cut is obtained by cutting the lower clique ($\{n,o,p,q,r,s\}$), yielding a minimum cut of size 2. The manipulator's utility is strictly improved from 5 to 6.
\end{proof}

\subsection{Max-Egal}
\label{Sec:egalitarian}
We now consider the objective of maximizing the egalitarian social welfare (Max-Egal), i.e., maximizing the utility of the agent that is worst off. Formally, it is $\underset{a\in A}{\min}(u(a,P))$.
The objective egalitarian social welfare has also been studied in ASHGs \cite{peters2016graphical,aziz2013computing}.
Maximizing the egalitarian SW might result in a decrease in the average utility of the agents  (which is correlated to the utilitarian SW) but it tries to ensure that all of the agents will have some minimum utility. Now, let $Eg(P,G)$ be the egalitarian SW of a coalition structure $P$ in graph $G$. The following theorems show that Max-Egal is resistant to manipulation by adding edges. The intuition is that by adding edges the manipulator is not able to pretend to be the agent with the minimum utility, and he may increase the utility of the other agents by at most $1$. 
\begin{theorem}
\label{thrm:egal_undir_add}
Max-Egal is weak-proof against manipulator $m^+$ over undirected networks.
\end{theorem}
\begin{proof}
Let 
$
u_0=\underset{P\in O(G)}{\min}(\{u(m,P)\}),
u_1=\underset{P\in O(G)}{\max}(\{u(m,P)\})
$.
We will refer to the CS yielding $u_0$ as $P_0$.
Assume by contradiction that Max-Egal is subject to weak-improvement.
That is, there exists a manipulation $r_{m}$ and a CS $P^m\in O(G^{m})$ such that $u(m,P^m) > u_1$. That is, $P^m\notin O(G)$. In addition, 
\begin{equation} \label{ineq:egal_undir}
\forall \  P \in O(G^m), u(m,P) > u_0.
\end{equation}

Since the manipulator can only add edges, it holds that $Eg(P^{m},G^m)\geq Eg(P_0,G)$. Moreover, if $Eg(P^{m},G^m)=Eg(P_0,G)$ then $P_{0}\in O(G^{m})$, which is not possible according to  inequality~\ref{ineq:egal_undir}. Therefore, $Eg(P^{m},G^m)>Eg(P_0,G)$. Since the manipulator is able to add at most one new edge to every agent and $G$ is undirected, then $\forall a \in A \setminus \{m\}$, \[
u(a,P^{m})  \geq Eg(P^{m},G^m) - 1 \geq Eg(P_0,G).
\]
In addition, $u(m,P^{m})>u_1\geq Eg(P_0,G)$. 
Overall, $\forall a \in A, u(a,P^{m}) \geq Eg(P_0,G)$. That is, $Eg(P^{m},G) \geq Eg(P_0,G)$, and thus $P^m\in O(G)$, which is a contradiction. 
\end{proof}

\begin{theorem}
\label{thrm:egal_dir_add}
Max-Egal is strategyproof against a manipulator $m^+$ over directed networks.
\end{theorem}
\begin{proof}
Let 
$
u_0=\underset{P\in O(G)}{\min}(\{u(m,P)\}),
u_1=\underset{P\in O(G)}{\max}(\{u(m,P)\})
$.
We will refer to the CS yielding $u_0$ as $P_0$. Note that for every $P\in O(G)$ it holds that $Eg(P,G) \leq u_0 $.
Assume by contradiction that Max-Egal is subject to UB-improvement.
That is, there exists a manipulation $r_{m}$ and a CS $P^m\in O(G^{m})$ such that $u(m,P^m) > u_1$. That is, $P^m\notin O(G)$.

Since the manipulator can only add edges, it holds that $Eg(P^{m},G^m)\geq Eg(P_0,G)$. Moreover, if $Eg(P^{m},G^m)=Eg(P_0,G)$ then $P_{0}\in O(G^{m})$, which is not possible. Therefore, $Eg(P^{m},G^m)>Eg(P_0,G)$. Recall that in directed networks the utility of the other agents does not change. Therefore $\forall a \in A \setminus \{m\}$, \[
u(a,P^{m})  \geq Eg(P^{m},G^m) > Eg(P_0,G).
\]
In addition, $u(m,P^{m})>u_1\geq Eg(P_0,G)$. 
Overall, $\forall a \in A, u(a,P^{m}) \geq Eg(P_0,G)$. That is, $Eg(P^{m},G) \geq Eg(P_0,G)$, and thus $P^m\in O(G)$, which is a contradiction.

Now, assume by contradiction that Max-Egal is subject to LB-improvement.
That is, there exists a manipulation $r_{m}$ such that
\begin{equation} \label{ineq:egal_dir}
\forall \  P \in O(G^m), u(m,P) > u_0.
\end{equation} That is $P_0\notin O(G^m)$.
Denote an arbitrary CS in $O(G^m)$ as $P^m$. 
It holds that $Eg(P^m,G^m)>Eg(P_0,G^m)$ and $Eg(P_0,G)\geq E(P^m,G)$.

Again, in directed networks the utility of the other agents does not change. Therefore, if after the manipulation $Eg(P^m,G^m)>Eg(P^m,G)$, it can only change by the utility of $m$. But $u(a,P^m)>u(a,P)$, hence even before the manipulation $Eg(P^m,G^m)>Eg(P^m,G)$, in contradiction.
\end{proof}

\subsection{At-least-1}
\label{sec:least1}
In the At-least-1 objective the organizer is only interested in ensuring that every agent will have a utility of at least $1$. This objective is very general, and it may result in many possible CSs. It has mostly been studied in the context of graph theory \cite{stiebitz1996decomposing,alon2006splitting,bang2016finding}.

Note that there are some instances where there is no CS that guarantees a utility of at least $1$ to every agent. We call such an instance \textit{infeasible}, and we then write $O(G) = \emptyset$. In infeasible instances we assume that the utility of all of the agents is $0$.
%
We show that, in contrast to the previous objectives, At-least-1 is less susceptible to manipulations. Specifically, we show that an UB-improvement is almost always impossible, and LB-improvement is impossible by adding directed edges. The intuition is that adding edges is beneficial only if the network is undirected and the new edges transform an infeasible instance into a feasible instance, and by removing edges the manipulator is not able to introduce new solutions.

\begin{theorem}
\label{thrm:least1_add_dir}
At-least-1 is strategyproof against manipulator $m^+$ over directed networks.
\end{theorem}
\begin{proof}
Let $u_1=\underset{P\in O(G)}{\max}(\{u(m,P)\})$, and recall that if $O(G) = \emptyset$ then $u_1=0$.
Assume by contradiction that the At-least-1 objective is subject to UB-improvement.
That is, there exists a manipulation $r_{m}$ and a coalition structure $P^m\in O(G^{m})$ such that $u(m,P^m) > u_1$. That is $P^m\notin O(G)$, and $\forall a \in A, |N^m(a)| \geq 1$.
Since $u_1 \geq 0$ then $u(m,P^m) \geq 1$. 
In addition, recall that in a directed network, $\forall a \in A \setminus \{m\}, N(a)=N^m(a)$, thus $\forall a \in A \setminus \{m\}, u(a,P^m) \geq 1$.
Overall, $\forall a \in A, u(a,P^m) \geq 1$ and thus $P^m \in O(G)$, which is a contradiction.



Regarding  LB-improvement, since the manipulator is only able to add edges then $O(G) \subseteq O(G^m)$. Therefore, no LB-improvement is possible if $O(G) \neq \emptyset$. 
If $O(G) = \emptyset$, then $u_1=0$. Now, assume by contradiction that the At-least-1 objective is subject to LB-improvement. That is, there exists a manipulation $r_{m}$ for which $\underset{P\in O(G^m)}{\min}(\{u(m,P)\})$ is at least $1$. Since $u_1=0$ that would imply an UB-improvement as well, which is impossible (as shown above).
\end{proof}

\begin{theorem}
\label{thrm:least1_remove}
At-least-1 is UB-proof against manipulator $m^-$ over directed and undirected networks.
\end{theorem}
\begin{proof}
Since $m$ is only able to remove edges then for every manipulation $r_m$ it holds that $O(G^m) \subseteq O(G)$. Therefore, $\underset{P\in O(G)}{\max}(\{u(m,P)\})\geq \underset{P\in O(G^m)}{\max}(\{u(m,P)\}) $ and no UB-improvement is possible. 
\end{proof}

\section{Limited Information}
\label{sec:limited}
We now focus on more realistic settings, in which the manipulator is not familiar with the full structure of the network. Instead, we assume that the manipulator is either familiar only with his immediate neighbours in the network, or he may also be familiar with the neighbours of his immediate neighbours. Within this setting we need to revise our definitions of successful manipulations. Specifically, since the manipulator is familiar only with a partial network, we define suitable safe manipulations. Informally, a safe manipulation is a manipulation in which the manipulator is not worse off in all of the possible completions of the partial network, and there exists at least one completion of the partial network in which the manipulator is better off.  

Formally, let $A_0 = \{m\}$.
Let $G_1=(A,E_1)$ be a graph, $E_1 \subseteq E$, where $(u,v) \in E_1$ if either $u$ or $v$ belongs to $A_0$.
Similarly, let $A_1 = \{u : (u,v) \in E_1 \lor  (v,u) \in E_1\}$.
Let $G_2=(A,E_2)$ be a graph, $E_2 \subseteq E$, where $(u,v) \in E_2$ if either $u$ or $v$ belongs to $A_1$.
%
%
%
Given $d \in \{1,2\}$, a \emph{possible network} $\overline{G_d}$ of $G_d$ is a network $\overline{G_d}=(A,\overline{E_d})$ where $\overline{E_d} = E_d \cup E'$ such that if $(u,v)\in E'$ then neither $u$ and $v$ belong to $A_{d-1}$.
We assume that the manipulator is familiar with $G_d$ and the objective $obj$.
We denote the settings in which the manipulator is familiar with $G_1$ ($G_2$) by \emph{distance 1} (\emph{distance 2}). 
Indeed, since the manipulator is always familiar with his immediate neighbours he can still add or remove edges as in the full information setting. However, since the manipulator is only familiar with $G_d$ he needs to consider the effect of his manipulation  on every possible network $\overline{G_d}$. Given $\overline{G_d}$ and a manipulation $r_m$, let $\overline{G_d}^m$ be the possible network after the manipulation $r_m$.
We can now revise our definitions of successful manipulations.
Given a partial network $G_d$ and an objective $obj$:
\begin{definition}
A manipulation $r_m$ is a \emph{d-safe lower bound improvement} for a manipulator $m$ if for all possible networks $\overline{G_d}$ of $G_d$ it holds that
\[
\underset{P\in O_{obj}(\overline{G_d}^m)}{\min}(u(m,P))\geq\underset{P\in O_{obj}(\overline{G_d})}{\min}(u(m,P)).
\]
and for at least one possible network $\overline{G_d}$, $r_m$ is a LB-improvement.



An objective is \emph{subject to d-safe LB-improvement} by manipulator $m$ over (un)directed networks if there exists a (un)directed partial network $G_d$ and a manipulation $r_m$ such that $r_m$ is a d-safe LB-improvement for $m$. Otherwise we say that \emph{obj} is \emph{d-safe LB-proof} against $m$.
\end{definition}
\noindent
The definitions for the other manipulation types are similar. Note that susceptibility to d-safe manipulation implies susceptibility to d'-safe manipulation for any $d'>d$, as well as to the full information case. Similarly, resistance to manipulation in the full information setting implies resistance to manipulation in the limited information setting.
Due to space constraints, the proofs of most of the theorems in this section are provided in the appendix. However, we  provide references to the figures that depict the safe manipulations.
\begin{figure*}[t]
    \centering
    \centering  
        \begin{subfigure}{0.1\textwidth}
            \centering
            \includegraphics[page=1,width=\textwidth]{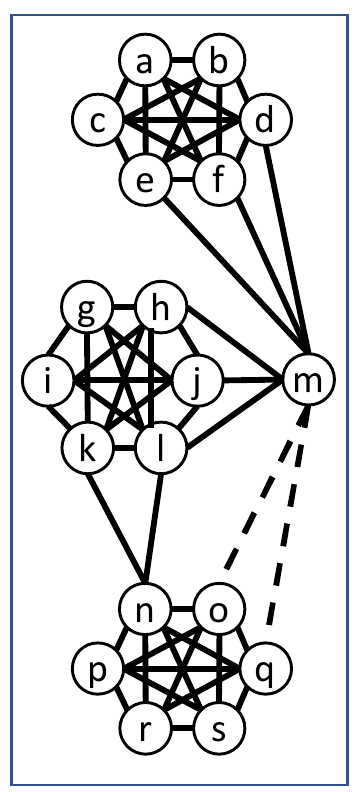}
            \caption{\\U/S}
            \label{fig:Util_remove_Improve}
        \end{subfigure}    
        \hfill
        \begin{subfigure}{0.1\textwidth}
            \centering
            \includegraphics[page=47,width=\textwidth]{Graphs/graphs.pdf}
            \caption{\\U/S}
            \label{fig:Util_dir_remove}
        \end{subfigure}
        \hfill
        \begin{subfigure}{0.1\textwidth}
        \centering
            \includegraphics[page=6,width=\textwidth]{Graphs/graphs.pdf}
            \caption{\\U/LB}
            \label{fig:Util_remove_LB}
        \end{subfigure}    
        \hfill
        \begin{subfigure}{0.1\textwidth}
        \centering
            \includegraphics[page=9,width=\textwidth]{Graphs/graphs.pdf}
            \caption{\\U/UB}
            \label{fig:Util_remove_UB}
        \end{subfigure}
        \hfill
        \begin{subfigure}{0.1\textwidth}
        \centering
            \includegraphics[page=19,width=\textwidth]{Graphs/graphs.pdf}
            \caption{\\E/S}
            \label{fig:Egal_directed_remove}
        \end{subfigure}
        \hfill
        \begin{subfigure}{0.1\textwidth}
            \centering
            \includegraphics[page=17,width=\textwidth]{Graphs/graphs.pdf}
            \caption{\\E/S}
            \label{fig:Egal_undirected_remove}
        \end{subfigure}
        \hfill
        \begin{subfigure}{0.1\textwidth}   \centering
            \includegraphics[page=31,width=\textwidth]{Graphs/graphs.pdf}
            \caption{\\1/LB}
            \label{fig:least1_remove}
        \end{subfigure}
    
    \caption{Figures for the proofs of manipulations by $m^-$. Key:U=Max-Util, E=Max-Egal, 1=At-least-1, S=strict-improvement, UB=UB-improvement, LB=LB-improvement}
    \label{fig:remove_graphs}
\end{figure*}

{\small
\begin{table}[t]
\ra{1}
\begin{tabular}{@{}llllr@{}}\toprule 
$Objective$ & $Manipulator$ & $Network$ &  $Type$ & $Fig$\\\midrule
Max-Util & Add & Both & Strict & \ref{fig:Util_add}\\
Max-Util & Remove & Directed & Strict & \ref{fig:Util_dir_remove}\\
Max-Util & Remove & Undirected & LB & \ref{fig:Util_remove_LB}\\
Max-Util & Remove & Undirected & UB & \ref{fig:Util_remove_UB}\\
Max-Egal & Add & Undirected & LB & \ref{fig:Egal_undirected_add_LB}\\
Max-Egal & Add & Undirected & UB & \ref{fig:Egal_undirected_add_UB}\\
Max-Egal & Remove & Directed & Strict & \ref{fig:Egal_directed_remove}\\
Max-Egal & Remove & Unirected & Strict & \ref{fig:Egal_undirected_remove}\\
At-Least-1 & Add & Undirected & Strict & \ref{fig:Least1_undirected_add}\\
At-Least-1 & Remove & Both & LB & \ref{fig:least1_remove}\\
\bottomrule
\end{tabular}
\caption{Summary of susceptibility results for distance 2.
Key: LB/UB/Strict = the objective is subject to LB/UB/Strict-improvement.
}
\label{tbl:distnce2_summary}
\end{table}
}
\subsection{Distance 1}
\label{sec:distance1}
We analyze the three objectives and their susceptibility to safe manipulations in the setting of distance 1. Remarkably, even in this setting there are situations where a safe manipulation exists (Proposition~\ref{prop:distance1}). 
However, Theorem~\ref{thm:all_distance1} shows that for most situations safe manipulation is impossible. The proof is based on extensive enumeration of networks, where we show that either no manipulation exists or there exists only an unsafe manipulation.

\begin{proposition}
\label{prop:distance1}
Max-Egal against manipulator $m^-$ over directed networks and At-Least-1 against $m^+$ over undirected networks are subject to 1-safe UB-improvement.
\end{proposition}

\begin{figure}[t]
    \centering
        \begin{subfigure}{0.09\textwidth}
            \centering
            \includegraphics[page=2,width=\textwidth]{Graphs/Distance_1_paper.pdf}
            \caption{R/U/LB}
            \label{fig:distance1_util_unsafe_remove_LB}
        \end{subfigure}
        \hfill
        \begin{subfigure}{0.09\textwidth}
            \centering
            \includegraphics[page=1,width=\textwidth]{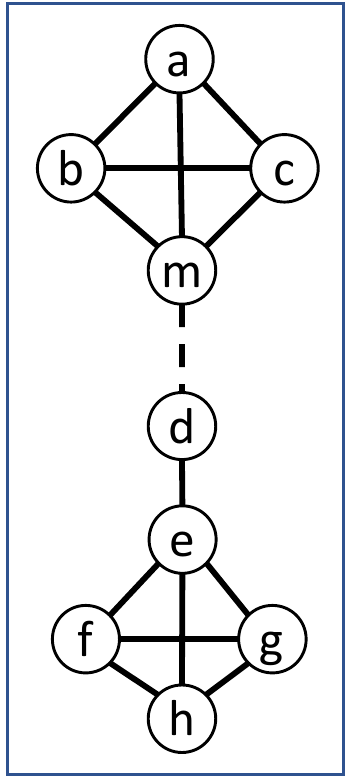}
            \caption{R/U/UB}
            \label{fig:distance1_util_unsafe_remove_UB}
        \end{subfigure}
                \hfill
                \begin{subfigure}{0.09\textwidth}
            \centering
            \includegraphics[page=4,width=\textwidth]{Graphs/Distance_1_paper.pdf}
            \caption{A/U/LB}
            \label{fig:distance1_util_unsafe_add_LB}
        \end{subfigure}
                \hfill
                \begin{subfigure}{0.09\textwidth}
            \centering
            \includegraphics[page=3,width=\textwidth]{Graphs/Distance_1_paper.pdf}
            \caption{A/U/UB}
            \label{fig:distance1_util_unsafe_add_UB}
        \end{subfigure}
                \hfill
                \begin{subfigure}{0.09\textwidth}
            \centering
            \includegraphics[page=5,width=\textwidth]{Graphs/Distance_1_paper.pdf}
            \caption{R/E/UB}
            \label{fig:distance1_egal_ub_any}
        \end{subfigure}
    
    \caption{Figures for the proofs of safe and unsafe manipulations, distance 1. Key:U=Max-Util, E=Max-Egal, UB=UB-improvement, LB=LB-improvement, R=Remove, A=Add}
    \label{fig:distance1_graphs}
\end{figure}

\begin{proof}
Figure~\ref{fig:distance1_egal_ub_any} provides proof for Max-Egal.
Clearly, At-Least-1 is subject to UB-improvement by simply adding any edge, as it can turn an infeasible instance into a feasible instance.
\end{proof}

\begin{theorem}
\label{thm:all_distance1}
All three objectives are 1-safe strategyproof except for the cases in Proposition~\ref{prop:distance1}.

\end{theorem}

\begin{proof}[Proof (partial)]
We show that Max-Util is 1-safe strategyproof against manipulators $m^-$ and $m^+$ over undirected networks. We start with $m^-$;
If $0<|N(m)|<n-1$, Figures \ref{fig:distance1_util_unsafe_remove_LB} and \ref{fig:distance1_util_unsafe_remove_UB} show that removing any edge is UB-unsafe and LB-unsafe, respectively.
Clearly, the graphs in these Figures can be extended to any number of agents and any number of neighbours of the manipulator. For example, to extend figure \ref{fig:distance1_util_unsafe_remove_LB} to arbitrary numbers of agents and neighbours, simply put all the manipulator and all of its neighbours but one, call him $a$ into a clique. Let the other agents form a clique of their own. Lastly connect $a$ to the other clique with only one edge. This way removing an edge is LB-unsafe.

If $|N(m)|=n-1$,
%
we first show that Max-Util is 1-safe UB-proof.
If the manipulator removes only 1 edge then he cannot improve his upper bound at all.
If the possible network was a complete graph with $n$ nodes then removing two edges or more is UB-unsafe.
To see that Max-Util is 1-safe LB-proof, look at a complete graph where one edge is missing (an edge not connected to $m$). Removing any edge in that case can only lower the manipulator's LB.

Continuing with $m^+$, if $0<|N(m)|<n-1$then Figures \ref{fig:distance1_util_unsafe_add_LB},\ref{fig:distance1_util_unsafe_add_UB} show that adding any edge is LB- and UB-unsafe respectively.
Again, these examples can be extended to fit any number of agents and any number of neighbours the manipulator has.
If $|N(m)|=n-1$ the manipulator cannot add edges.
The proof for the other settings is deferred to the appendix. 
\end{proof}

\subsection{Distance 2}
\label{subsecdist-2}
Unlike in the distance 1 setting, the results for the distance 2 setting are almost the same as the results in the full information setting. Indeed, all of the resistance results are derived by the resistance results in the full information setting. Therefore, in this section we provide only  susceptibility results summarized in Table~\ref{tbl:distnce2_summary} where each entry represents a situation, what type of manipulation it is subject to, and reference to a figure providing a proof. Note that the figures showcase a possible network. The partial network known to the manipulator can easily be derived from them. 
Overall, we show that, surprisingly, all of the results for the full information setting hold for the distance 2 setting, except for one case: when maximizing the utilitarian SW, with a $m^-$ manipulator over undirected networks. In this case, Proposition~\ref{Prop:util_full_info} shows that the objective is subject to strict-improvement with full information while Figures~\ref{fig:Util_remove_LB} and \ref{fig:Util_remove_UB} only show that it is subject to 2-safe LB- and UB-improvement in the distance 2 setting. Indeed, we believe that the objective is 2-safe weak-proof.

\section{Discussion}
In this section we discuss our results. We explain the different phenomena that we observe when comparing the results for the different settings.
Our results indicate that manipulations over undirected networks are easier than manipulations over directed networks. This is due to the fact that in directed networks the manipulator can only influence his own utility. In undirected networks the manipulator can actually influence the utility of his neighbours, and this additional power enables the manipulation in additional situations.
We can also observe that Max-Util is the easiest objective to manipulate when compared with Max-Egal and At-least-1. Indeed, Max-Util is inherently different from the other two objectives: when maximizing the utilitarian SW, the organizer is interested in the average utility, thus the organizer always takes into account the utility of the manipulator. In contrast, in the other two objectives the organizer is interested in the utility of the weakest agents, thus the organizer may not take into account the utility of the manipulator (e.g., if the manipulator already has a utility of $5$ in the At-least-1 objective). 
This characteristic of Max-Egal and At-least-1 objectives can also explain why it is easier to manipulate these objectives by removing edges rather than by adding edges: by removing edges the manipulator can pretend to be the weakest agent, affecting the organizer's choice of the coalition structure. Following this observation, we would expect that manipulating the At-least-1 objective will be harder than manipulating the Max-Egal objective: in At-least-1 the organizer is interested only in ensuring a minimum utility of $1$ while in Max-Egal the organizer is also interested in maximizing the minimal utility. However, it turns out that At-least-1 can be manipulated by turning an infeasible instance into a feasible instance, thus there are situations in which At-least-1 is subject to strict-improvement while Max-Egal is weak-proof.
Finally, we note that there is a significant difference between results for the settings of distance 1 and distance 2. In the setting of distance 1, the manipulator has very limited knowledge of the network. Therefore, it is hard for the manipulator to estimate the full effect of adding or removing edges, thus finding that a safe manipulation is impossible in most of the situations. Surprisingly, even though the manipulator is not familiar with the full structure of the network in the setting of distance 2, the additional information in this setting is sufficient for finding safe manipulations in many situations.



\section{Conclusions and Future Work}
We have studied manipulation in the setting of a central organizer that would like to partition a social network into $k$ coalitions. 
The organizer has a certain objective she would like to satisfy or maximize, whereas there is a manipulator that would like to maximize his utility, i.e., the number of friends within his coalition.
We have distinguished between a manipulator who has full information regarding the structure of the social network, and a manipulator that is only familiar with the edges in close proximity to him. 
An important future research direction to explore is the complexity of finding a manipulation, given a specific objective. We would also like to extend our analysis to the setting with weighted social networks, or social networks with negative edges.


\clearpage
\bibliographystyle{named}
\bibliography{bib}

\end{document}


\maketitle

\clearpage
\section*{Appendix}

\subsection*{Supplementary proofs}
In the appendix we use the term \emph{LB- (UB)-unsafe} to refer to a manipulation that lowers the lower (upper) bound of the manipulator. Also, recall that a \emph{solution} refers to any CS that satisfies the objective.

\subsection*{Proof of Theorem~\ref{thrm:egal_dir_add}}
Let 
$
u_0=\underset{P\in O(G)}{\min}(\{u(m,P)\}),
u_1=\underset{P\in O(G)}{\max}(\{u(m,P)\})
$.
We will refer to the CS yielding $u_0$ as $P_0$. Note that for every $P\in O(G)$ it holds that $Eg(P,G) \leq u_0 $
Assume by contradiction that Max-Egal is subject to UB-improvement.
That is, there exists a manipulation $r_{m}$ and a CS $P^m\in O(G^{m})$ such that $u(m,P^m) > u_1$. That is, $P^m\notin O(G)$.

Since the manipulator can only add edges it holds that $Eg(P^{m},G^m)\geq Eg(P_0,G)$. Moreover, if $Eg(P^{m},G^m)=Eg(P_0,G)$ then $P_{0}\in O(G^{m})$, which is not possible. Therefore, $Eg(P^{m},G^m)>Eg(P_0,G)$. Recall that in directed networks the utility of the other agents does not change. Therefore $\forall a \in A \setminus \{m\}$, \[
u(a,P^{m})  \geq Eg(P^{m},G^m) > Eg(P_0,G).
\]
In addition, $u(m,P^{m})>u_1\geq Eg(P_0,G)$. 
Overall, $\forall a \in A, u(a,P^{m}) \geq Eg(P_0,G)$. That is, $Eg(P^{m},G) \geq Eg(P_0,G)$, and thus $P^m\in O(G)$, which is a contradiction.

Now, assume by contradiction that Max-Egal is subject to LB-improvement.
That is, there exists a manipulation $r_{m}$ such that
\begin{equation} \label{ineq:egal_dir}
\forall \  P \in O(G^m), u(m,P) > u_0.
\end{equation} That is $P_0\notin O(G^m)$.
Denote an arbitrary CS in $O(G^m)$ as $P^m$. 
It holds that $Eg(P^m,G^m)>Eg(P_0,G^m)$ and $Eg(P_0,G)\geq E(P^m,G)$.

Again, in directed networks the utility of the other agents does not change. Therefore if after the manipulation $Eg(P^m,G^m)>Eg(P^m,G)$ it can only change by the utility of $m$. But $u(a,P^m)>u(a,P)$, hence even before the manipulation $Eg(P^m,G^m)>Eg(P^m,G)$, in contradiction.

\subsection*{Proof of Theorem~\ref{thm:all_distance1}}

\subsubsection*{Max-Util}
If $0<|N(m)|<n-1$, by looking at possible networks like the ones in Figures \ref{fig:distance1_util_unsafe_add_LB},\ref{fig:distance1_util_unsafe_add_UB} we can see that adding any edge is LB- and UB-unsafe respectively.
These examples can be tweaked to fit any number of agents and any number of desired agents the manipulator has.
If $|N(m)|=n-1$ the manipulator cannot add edges.

\subsubsection*{Max-Egal}
Obviously adding or removing in undirected networks is both LB- and UB-unsafe. Adding an edge towards agent $a_0$ can lead to utility $0$ when the maximum egalitarian SW is $1$, if $m$ ends up in a coalition alone with $a_0$.
Removing an edge can lead to utility $0$. If one $m$'s neighbours has no other agents, removing that edge results in maximum egalitarian SW $0$, hence the lower bound is also $0$. So both add and remove are LB-unsafe.
For the cases of $|N(m)|<n-3, |N(m)|=n-2$ and $|N(m)|=n-1$ see Figures~\ref{fig:egalremove1},\ref{fig:egalremove2} and \ref{fig:egalremove3} for proofs that removing is UB-unsafe, respectively. If $m$ has more neighbours can just add nodes on the top, and more non-neighbours in bottom of the figures.
Figure~\ref{fig:egalremove4} provides proof that adding UB-unsafe. For higher numbers both neighbours and non-neighbours are added at the bottom. Note that this example requires at least neighbours ($|N(m)|\geq2$. However, if $m$ has only one neighbour it is easy to see no manipulation is beneficial against Max-Egal.

\subsubsection*{At-Least-1}
Adding any edge to another agent $a_0$ might result in a LB- of $0$. That is because the CS where $a_0$ and $m$ form one coalition could be a solution now.
Any partial network in distance 1 can have a possible network with a solution where $m$'s original LB- is higher than $0$, hence adding is LB-unsafe.


For At-Least-1, all that is left to prove is that against manipulator $m^-$ the objective is LB-proof. Of course, over undirected networks it is true, as any removal might lead to an infeasible instance.

Over directed networks, it is also easy to see the any removal may lead to an infeasible instance; If $m$ has at least one neighbour $a_0$ for which both $(a_0,m),(m,a_0)\in E $, it could be the case that the only solution is where $m$ and $a_0$ form a coalition alone. (If all of the other agents form a directed circle, and have no other edges but ones going to or from $m$). In that case removing is LB-unsafe.
If $m$ does not have a neighbour like that, the same example still applies when the only solution is $m$ alongside one of his out-neighbours and another agent. i.e. $a_0,a_1$ where $(m,a_0),(a_0,a_1),(a_1,a_0)\in E$.

\clearpage
\subsection*{Equal Size}
An important constraint that arises in many real world scenarios is that all of the coalitions should be of the same size. For example, when dividing students into classes it is common to ask that all of the classes consist of roughly the same number of students.
Therefore we also consider the setting in which all of the coalitions are of the size $n/k$, if $n/k$ is an integer. Otherwise, some coalitions are of size $\left \lceil{n/k}\right \rceil$ and some are of size $\left \lfloor{n/k}\right \rfloor$. We call this restriction the Equal Size constraint, and through the paper we analyze the different objectives with or without it.

Due to space constraints, the definition and results of the equal size constraints have been moved to the appendix. All of the resistance results from Section~\ref{sec:full_info} still stand with the equal size constraint. The susceptibility results are summarized in Table~\ref{tbl:distnce2_equalsize}.

\begin{table}[htp]

\ra{1}
\begin{tabular}{@{}lcccr@{}}\toprule 
$Objective$ & $M$ & $Network$ & $Type$ & $Figure$\\\midrule
Max-Util & A & Both & Strict & \ref{fig:util_size_add}\\
Max-Util & R & Both & Strict & \ref{fig:Util_size_remove}\\
Max-Egal & A & Undirected & LB & \ref{fig:Egal_size_undirected_add_LB}\\
Max-Egal & A & Undirected & UB & \ref{fig:Egal_size_undirected_add_UB}\\
At-Least-1 & A & Undirected & Strict & \ref{fig:Least1_size_undirected_add}\\
At-Least-1 & R & Both & LB & \ref{fig:least1_size_remove}\\
\bottomrule
\end{tabular}
\caption{Summary of susceptibility results for distance 2 with the equal size constraint.
Key: A = add, R = Remove, LB/UB/Strict = the objective is subject to LB/UB/Strict-improvement.
}
\label{tbl:distnce2_equalsize}
\end{table}

\begin{proposition}
\label{thm:util_distance1}
Max-Util with the equal size constraint is subject to 1-safe LB- and UB-improvement and is 1-safe weak-proof for a manipulator $m^-$ over directed and undirected networks. 
\end{proposition}

\begin{proof}
See Figures~\ref{fig:distance1_util/egal_lb} and \ref{fig:distance1_util_ub} for the 1-safe LB-improvement and 1-safe UB-improvement, respectively. We move on to prove it is 1-safe weak-proof.

First, we prove that if the manipulator has less than $n-1$ neighbours, no UB-safe manipulation exists.
If $0<|N(m)|\leq \lfloor n/2 \rfloor$ then Figures~\ref{fig:thrm5_1} and \ref{fig:thrm5_2} provide examples for UB-unsafe scenarios for even and odd $n$, respectively.
If $\lfloor n/2 \rfloor<|N(m)| < n/2$ then Figures~\ref{fig:thrm5_3} and \ref{fig:thrm5_4} provide examples for UB-unsafe scenarios for even and odd $n$, respectively.

If $|N(m)| = n-1$, then if $n$ is even, the manipulator always gets the same utility and no manipulation is possible.
If $n$ is odd we have shown it is subject to LB and UB 1-safe improvement.
If the manipulator removes more than 1 edge, Figure~\ref{fig:thrm5_5} provides an example that this is UB-unsafe.

By removing only 1 edge, the manipulation cannot be weak-improvement. 
Denote the utilitarian SW of coalition structure $P$ in $G$ as $Ut(P,G)$.
Let 
$
u_0=\underset{P\in O(G)}{\min}(\{u(m,P)\}),
u_1=\underset{P\in O(G)}{\max}(\{u(m,P)\})
$.
We will refer to the CS yielding $u_0$ as $P_0$.
Assume by contradiction that Max-Util is subject to weak-improvement when removing only 1 edge.
That is, there exists a manipulation $r_{m}$ and a CS $P^m\in O(G^{m})$ such that $u(m,P^m) > u_1$. That is, $P^m\notin O(G)$. Also, $P_0\notin O(G^m)$.

Therefore, we can see that $Ut(P_0,G) > Ut(P^m,G)$ and $Ut(P^m,G^m)>Ut(P_0,G^m)$.
Since the manipulator is only able to remove edges it holds that $Ut(P^m,G^m)\leq Ut(P^m,G)$. So we get that $Ut(P_0,G) > Ut(P^m,G^m) \xrightarrow{} Ut(P_0,G) - 1 \geq Ut(P^m,G^m)$
Since only 1 edge was removed, the utilitarian SW could drop at most by $1$ and it holds that $Ut(P^0,G) - 1 \leq Ut(P^0,G^m)$.
Combining the last two inequalities we get $ Ut(P_0,G^m) \geq Ut(P_m,G^m) $ in contradiction.
\end{proof}

\begin{proposition}
\label{thm:Egal_distance1}
Max-Egal with the equal size constraint is subject to 1-safe LB-improvement and is 1-safe UB-proof for a manipulator $m^-$ over directed networks.
\end{proposition}

\begin{proof}
See Figure~\ref{fig:distance1_util/egal_lb} for the 1-safe LB-improvement.

For the cases of $|N(m)|<n-3, |N(m)|=n-2$ and $|N(m)|=n-1$ Figures~\ref{fig:egalremove1},\ref{fig:egalremove2} and \ref{fig:egalremove3} show the that removing any edge is UB-unsafe, respectively.
\end{proof}

\begin{proposition}
\label{thm:least1_distance1_es}
At-Least-1 with the equal size constraint is subject to 1-safe UB-improvement and is 1-safe LB-proof for a manipulator $m^+$ over undirected networks.
\end{proposition}

\begin{proof}
See Figure~\ref{fig:distance1_least1} for the 1-safe UB-improvement.
The resistance proof is the same as without the equal size constraint.
\end{proof}

\begin{theorem}
\label{thm:all_distance1_es}
Except for the situations in Theorems~\ref{thm:util_distance1},\ref{thm:Egal_distance1} and \ref{thm:least1_distance1_es}, all of our objectives with the equal size constraint are 1-safe strategyproof. 
\end{theorem}
Here we separate the proof between the objectives as well:
\subsubsection*{Max-Util}
To show the manipulation is LB-unsafe;
If $0<|N(m)|\leq \lfloor n/2 \rfloor$ then Figures~\ref{fig:unsafeutil3} and \ref{fig:unsafeutil4} provide examples for even and odd $n$, respectively.
If $\lfloor n/2 \rfloor<|N(m)| < n/2$ then Figures~\ref{fig:unsafeutil7} and \ref{fig:unsafeutil8} provide examples for even and odd $n$, respectively.

To show the manipulation is UB-unsafe;
If $0<|N(m)|\leq \lfloor n/2 \rfloor$ then Figures~\ref{fig:unsafeutil1} and \ref{fig:unsafeutil2} provide examples for even and odd $n$, respectively.
If $\lfloor n/2 \rfloor<|N(m)| < n/2$ then Figures~\ref{fig:unsafeutil5} and \ref{fig:unsafeutil6} provide examples for even and odd $n$, respectively.

\subsubsection*{Max-Egal}
Manipulator $m^-$:
If $|N(m)| \leq n/2 -1$, a possible network which is an $n-1$ clique except for $m$ yields an egalitarian SW of exactly $|N(m)|$. Removing any of her neighbours is LB-unsafe as she is guaranteed to get all of the neighbours she reports.
Now look at a possible network where $|N(m)|-1$ neighbours of $m$ are connected only to her and all non-neighbours of $m$ are connected to another node $a_0$. Out of the non-neighbours, $n/2 -2 $ nodes have no more edges. The others are neighbours to neighbours of $m$. The last neighbour of $m$ is connected also to $a_0$. Of course, the maximum possible egalitarian SW is $1$, and the two possible outcomes for $m$ are either getting all the neighbours, or not getting the single neighbours which is connected to $a_0$. By removing an edge it might be the case that this single neighbour is the one being disconnected, hence removing an edge is UB-unsafe.
If $n/2 -1 < |N(m) < n-1$, look at a possible network where $n/2-1$ of neighbours of $m$ form a path $a_1, a_2, .. ,a_{n/2-1}$. All the other nodes form a clique of size $n/2$ and one of them $a_0$ is connected to $a_1$. The highest egalitarian SW achievable is $2$, by putting $m$ alongside the path. By removing the edge towards $a_1$ or $a_2$ the highest egalitarian SW possible is now 1, and it can also be achieved by swapping $m$ and $a_0$. Hence removing is LB-unsafe.
In a possible network where the $n/2-1$ neighbours are connected only to $m$ and $a_0$ (but not between themselves) the highest egalitarian SW possible is $1$, either by $m$ or $a_0$ being with the $n/2-1$ neighbours. By removing an edge towards one of the neighbours, $m$ is guaranteed to not be with them, but in the other coalition with only one neighbour, hence removing is UB-unsafe.
If $|N(m)=n-1$, if $n$ is even no manipulation is beneficial as the manipulator always gets her highest utility. 
If $n$ is odd, the possible network could be made of two cliques $L_1$ and $L_2$ of sizes $n-1/2$ and $n-1/2 -1$, and $L_1$ is missing 1 edge between two nodes $a_1, a_2$. The last node $a_0$ is connected to the whole network just like $m$. The two possible CSs yielding an egalitarian SW of $n-1/2 -1$ are where $L_1$ form a coalition alongside either $m$ or $a_0$ and $L_2$ is with the other. Hence the upper bound for $m$ is $n-1/2$. By removing an edge towards either $a_1$ or $a_2$, $m$ is guaranteed to be with $L_2$, so removing is UB-unsafe.

Manipulator $m^+$:
If $|N(m)| \leq n/2 -1 $, look at a network that is made of a star of size $n/2$ that includes all of $m$'s neighbour, another star of size $n/2-2$, and a node $a_0$ with no edges. Adding an edge towards $a_0$ is guaranteeing utility $0$ for the manipulator, as him being  being with $a_0$ and the smaller star is the only CS yielding egalitarian SW of $1$. Hence it is both LB and UB-unsafe.
If $n/2-1 < |N(m)| < n-1$, it is possible the graph is a clique guaranteeing $m$ a full coalition of neighbours. By adding any edge this might make some of the neighbours fake, therefore UB-unsafe. Lastly, look at a graph where $n/2 -1$ neighbours of $m$ form a circle $C_m$. Also, another $n/2-1$ nodes form a circle $C_0$, and call the last node $a_0$. $a_0$ is connected to two nodes $C_0$ and to one in $C_m$. Also, there is a node $a_1$ in $C_m$ that is connected to two nodes in $C_0$. Adding an edge to $a_0$ is LB-unsafe, as now a CS where $m$ is with $a_0$ instead of $a_1$ is a solution.

\subsubsection*{At-Leas-1}
If $|N(m)|<\lfloor n/2 \rfloor$, removing can still lead to an infeasible instance. If $|N(m)|<\lfloor n/2 \rfloor$ then removing edges while still having more than $\lfloor n/2 \rfloor$ does not change the outcome, as any previous solution is still satisfying At-Least-1. This is because the manipulator is still guaranteed at least $1$ out neighbour in every CS, and the other agents' utility has not changed. Removing edge until having less than $\lfloor n/2 \rfloor$ can again result in infeasible instance.

\begin{figure*}[t]
    \centering
    \begin{subfigure}{0.6\textwidth}
    \centering  
        \begin{subfigure}{0.15\textwidth}
        \renewcommand\thesubfigure{\alph{subfigure}1}
            \centering
        \includegraphics[page=5,width=\textwidth]{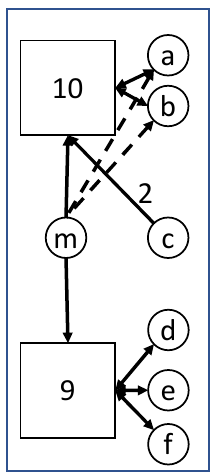}
        \caption{\\U/S/ES}
        \label{fig:util_size_add}
        \end{subfigure}
        \hfill
        \begin{subfigure}{0.15\textwidth}
            \addtocounter{subfigure}{-1}
            \renewcommand\thesubfigure{\alph{subfigure}2}
            \centering
        \includegraphics[page=23,width=\textwidth]{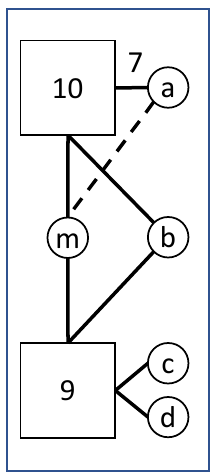}
        \caption{\\E/LB/ES}
        \label{fig:Egal_size_undirected_add_LB}
        \end{subfigure}
        \hfill
        \begin{subfigure}{0.15\textwidth}
            \addtocounter{subfigure}{-1}
            \renewcommand\thesubfigure{\alph{subfigure}3}
            \centering
        \includegraphics[page=25,width=\textwidth]{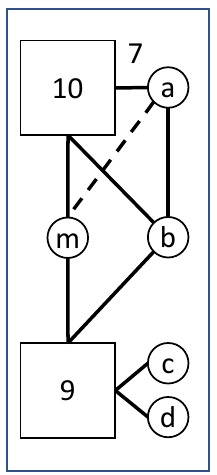}
        \caption{\\E/UB/ES}
        \label{fig:Egal_size_undirected_add_UB}
        \end{subfigure}
        \hfill
        \begin{subfigure}{0.15\textwidth}
                    \addtocounter{subfigure}{-1}
            \renewcommand\thesubfigure{\alph{subfigure}4}
            \centering
        \includegraphics[page=33,width=\textwidth]{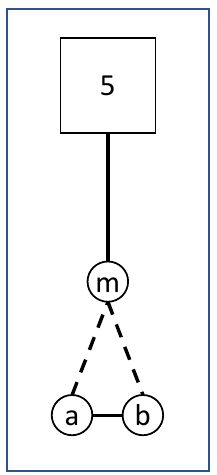}
        \caption{\\1/S/ES}
        \label{fig:Least1_size_undirected_add}
        \end{subfigure}
    \addtocounter{subfigure}{-1}
    \caption{$m^+$}
    \label{fig:add_es_subfig} 
    \end{subfigure}
    \hfill
    \begin{subfigure}{0.3\textwidth}
    \centering  
        \begin{subfigure}{0.3\textwidth}
            \renewcommand\thesubfigure{\alph{subfigure}1}
            \centering
        \includegraphics[page=45,width=\textwidth]{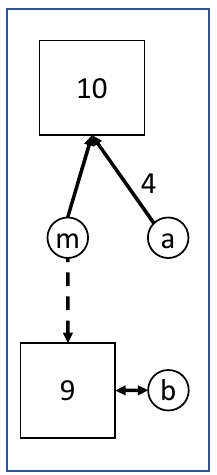}
        \caption{\\U/S/ES}
        \label{fig:Util_size_remove}
        \end{subfigure}    
        \hfill
        \begin{subfigure}{0.3\textwidth}
        \addtocounter{subfigure}{-1}
        \renewcommand\thesubfigure{\alph{subfigure}2}
            \centering
        \includegraphics[page=21,width=\textwidth]{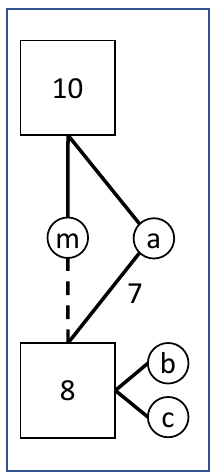}
        \caption{\\E/S/ES}
        \label{fig:Egal_size_remove}
        \end{subfigure}
        \hfill
        \begin{subfigure}{0.3\textwidth}
                    \addtocounter{subfigure}{-1}
        \renewcommand\thesubfigure{\alph{subfigure}3}
        \centering
        \includegraphics[page=27,width=\textwidth]{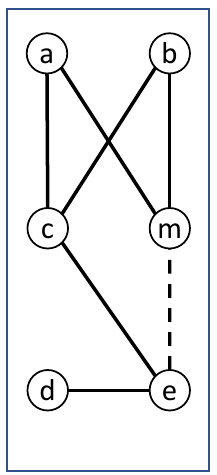}
        \caption{\\1/LB/ES}
        \label{fig:least1_size_remove}
        \end{subfigure}    
        \addtocounter{subfigure}{-1}
    \caption{$m^-$}
    \label{fig:remove_es_subfig} 
    \end{subfigure}


    \caption{Figures providing examples of susceptibility to manipulation with the equal size constraint. Key: U=Max-Util, E=Max-Egal, 1=At-least-1, S=strict-improvement, UB=UB-improvement, LB=LB-improvement}
    \label{fig:es_graphs}
\end{figure*}

\begin{figure}
    \centering
    \begin{subfigure}{0.07\textwidth}
        \centering
        \includegraphics[page=1,width=\textwidth]{Graphs/Distance 1 special cases.pdf}
        \caption{R,UB}
        \label{fig:thrm5_1}
    \end{subfigure}
    \hfill
    \begin{subfigure}{0.07\textwidth}
        \centering
        \includegraphics[page=2,width=\textwidth]{Graphs/Distance 1 special cases.pdf}
        \caption{R,UB}
        \label{fig:thrm5_2}
    \end{subfigure}
    \hfill
    \begin{subfigure}{0.07\textwidth}
        \centering
        \includegraphics[page=3,width=\textwidth]{Graphs/Distance 1 special cases.pdf}
        \caption{R,UB}
        \label{fig:thrm5_3}
    \end{subfigure}
    \hfill
    \begin{subfigure}{0.07\textwidth}
        \centering
        \includegraphics[page=4,width=\textwidth]{Graphs/Distance 1 special cases.pdf}
        \caption{R,UB}
        \label{fig:thrm5_4}
    \end{subfigure}
    \hfill
    \begin{subfigure}{0.07\textwidth}
        \centering
        \includegraphics[page=5,width=\textwidth]{Graphs/Distance 1 special cases.pdf}
        \caption{R,UB}
        \label{fig:thrm5_5}
    \end{subfigure}
    \\
    \begin{subfigure}{0.07\textwidth}
        \centering
        \includegraphics[page=6,width=\textwidth]{Graphs/Distance 1 special cases.pdf}
        \caption{A,UB}
        \label{fig:unsafeutil1}
    \end{subfigure}
    \hfill
    \begin{subfigure}{0.07\textwidth}
        \centering
        \includegraphics[page=7,width=\textwidth]{Graphs/Distance 1 special cases.pdf}
        \caption{A,UB}
        \label{fig:unsafeutil2}
    \end{subfigure}
    \hfill
    \begin{subfigure}{0.07\textwidth}
        \centering
        \includegraphics[page=8,width=\textwidth]{Graphs/Distance 1 special cases.pdf}
        \caption{A,LB}
        \label{fig:unsafeutil3}
    \end{subfigure}
    \hfill
    \begin{subfigure}{0.07\textwidth}
        \centering
        \includegraphics[page=9,width=\textwidth]{Graphs/Distance 1 special cases.pdf}
        \caption{A,LB}
        \label{fig:unsafeutil4}
    \end{subfigure}
    \\
    \begin{subfigure}{0.07\textwidth}
        \centering
        \includegraphics[page=10,width=\textwidth]{Graphs/Distance 1 special cases.pdf}
        \caption{A,UB}
        \label{fig:unsafeutil5}
    \end{subfigure}
    \hfill
    \begin{subfigure}{0.07\textwidth}
        \centering
        \includegraphics[page=11,width=\textwidth]{Graphs/Distance 1 special cases.pdf}
        \caption{A,UB}
        \label{fig:unsafeutil6}
    \end{subfigure}
    \hfill
    \begin{subfigure}{0.07\textwidth}
        \centering
        \includegraphics[page=12,width=\textwidth]{Graphs/Distance 1 special cases.pdf}
        \caption{A,LB}
        \label{fig:unsafeutil7}
    \end{subfigure}
    \hfill
    \begin{subfigure}{0.07\textwidth}
        \centering
        \includegraphics[page=13,width=\textwidth]{Graphs/Distance 1 special cases.pdf}
        \caption{A,LB}
        \label{fig:unsafeutil8}
    \end{subfigure}
    \caption{Unsafe manipulations, Max-Util, with the equal size constraint. Legend: A - add, R - remove, LB/UB - LB/UB-unsafe. In the figures $n$ stands for the total number of agents and $x$ for any number between $0$ and the clique's size}
    \label{fig:unsafe_util_equalsize}
\end{figure}

\begin{figure}
    \centering
\begin{subfigure}{0.07\textwidth}
        \centering
        \includegraphics[page=1,width=\textwidth]{Graphs/Distance 1 util unsafe.pdf}
        \caption{R/LB}
        \label{fig:distance1_util_unsafe_remove_LB_appendix}
    \end{subfigure}
    \hfill
    \begin{subfigure}{0.07\textwidth}
        \centering
        \includegraphics[page=2,width=\textwidth]{Graphs/Distance 1 util unsafe.pdf}
        \caption{R/UB}
        \label{fig:distance1_util_unsafe_remove_UB_appendix}
    \end{subfigure}
    \hfill
    \begin{subfigure}{0.07\textwidth}
        \centering
        \includegraphics[page=3,width=\textwidth]{Graphs/Distance 1 util unsafe.pdf}
        \caption{A/LB}
        \label{fig:distance1_util_unsafe_add_LB}
    \end{subfigure}
    \hfill
    \begin{subfigure}{0.07\textwidth}
        \centering
        \includegraphics[page=4,width=\textwidth]{Graphs/Distance 1 util unsafe.pdf}
        \caption{A/UB}
        \label{fig:distance1_util_unsafe_add_UB}
    \end{subfigure}

    \caption{Unsafe manipulations, Max-Util. Legend: A - add, R - remove, LB/UB - LB/UB-unsafe}
    \label{fig:distance1_unsafe_util_appendix}
\end{figure}

\begin{figure}
    \centering
    \begin{subfigure}{0.07\textwidth}
        \centering
        \includegraphics[page=1,width=\textwidth]{Graphs/egal unsafe distance1.pdf}
        \caption{R/UB}
        \label{fig:egalremove1}
    \end{subfigure}
    \hfill
    \begin{subfigure}{0.07\textwidth}
        \centering
        \includegraphics[page=2,width=\textwidth]{Graphs/egal unsafe distance1.pdf}
        \caption{R/UB}
        \label{fig:egalremove2}
    \end{subfigure}
    \hfill
    \begin{subfigure}{0.07\textwidth}
        \centering
        \includegraphics[page=3,width=\textwidth]{Graphs/egal unsafe distance1.pdf}
        \caption{R/UB}
        \label{fig:egalremove3}
    \end{subfigure}
    \hfill
    \begin{subfigure}{0.07\textwidth}
        \centering
        \includegraphics[page=4,width=\textwidth]{Graphs/egal unsafe distance1.pdf}
        \caption{A/UB}
        \label{fig:egalremove4}
    \end{subfigure}

    \caption{Unsafe manipulations, Max-Egal. Legend: A - add, R - remove, LB/UB - LB/UB-unsafe}
    \label{fig:distance1_unsafe_egal}
\end{figure}

\begin{figure}
    \centering
    \begin{subfigure}{0.07\textwidth}
        \centering
        \includegraphics[page=2,width=\textwidth]{Graphs/Distance 1 manipulation.pdf}
        \caption{\\U,E/LB}
        \label{fig:distance1_util/egal_lb}
    \end{subfigure} \hspace{20pt}
    \begin{subfigure}{0.07\textwidth}
        \centering
        \includegraphics[page=4,width=\textwidth]{Graphs/Distance 1 manipulation.pdf}
        \caption{\\U/UB}
        \label{fig:distance1_util_ub}
    \end{subfigure} \hspace{20pt}
    \begin{subfigure}{0.07\textwidth}
        \centering
        \includegraphics[page=7,width=\textwidth]{Graphs/Distance 1 manipulation.pdf}
        \caption{\\1/UB}
        \label{fig:distance1_least1}
    \end{subfigure}
    \caption{Manipulations by removing edges, for distance 1. Legend: Objective/Manipulation~Type. Key: U=Max-Util, E=Max-Egal, 1=At-Least-1, UB=UB-improvement, LB=LB-improvement.}
    \label{fig:distance1_manipulation}
\end{figure}

\clearpage

\subsection*{Max-Util Distance 2 Conjecture}
\begin{conjecture}
\label{conj:util}
Max-Util is 2-safe weak-proof against manipulator $m^-$ over undirected networks.
\end{conjecture}

We lay out some of the advances we have made trying to prove the conjecture:

Let $G_2=(A,E)$ be a partial network and $m^-$ a manipulator. Assume by negation a 2-safe weak-improvement $r^m$ exists over $G_2$. Hence there exists a possible game $\overline{G}$ of $G_2$ for which $r^m$ is a weak-improvement. We want to show that either the manipulation does not really improve the upper bound, or there exists another supplement $\overline{G'}$ of $G$ for which $r^m$ lowers the manipulator's upper bound.

Denote $\overline{G}^m$ as the network $\overline{G}$ after the manipulation $r^m$. Let $P^m=\{B_1,B_2\}$ be (one of) the CSs with higher utility for $m$ than all CSs in $\overline{G}$, and denote by $B_1$ the coalition in which $m$ is. Formally:
\begin{equation}
\label{ineq:util_ub_safe}
u(m,P_m) > \underset{P\in O_{obj}(\overline{G})}{\min}(u(m,P)).    
\end{equation}
Denote by $c_0$ and $c_m$ the minimum cut size of $\overline{G}$ and $\overline{G}^m$ respectively.
We will define $c(P,G)$ as the cut size of coalition structure $P$ in network $G$.

\begin{lemma}
$B_2$ contains at least $2$ node $a_1,a_2$ such that $a_1,a_2\in N$ and $a_1,a_2\notin N^m$.
\end{lemma}
\begin{proof}
Since $r^m$ is a LB-improvement there is a solution for $\overline{G}$ which is not a solution in $\overline{G}^m$. Since we only removed edges, the only reason it would happen is that $c_0 > c_m$.
Since $P^m$ is not a solution in $\overline{G}$ (as it is better than all solutions in it) we get that $c(P^m,\overline{G}) > c_0$. Combining both outcomes we get that $c(P^m,\overline{G}) \geq c_m+2$. Because we know that $c(P^m,\overline{G}^m) = c_m$ and the only changes between $\overline{G}$ and $\overline{G}^m$ are edges that the manipulator removed, it has to be the case that he removed at least $2$ edges going out to $B_2$.
\end{proof}
We get that the maximum utility $m$ can have after the manipulation is $|N(m)|-2$. Because of that, the minimum degree $\delta(\overline{G}) = min(\{deg(v) | v\in V\ and v != m \})$ of all the nodes but $m$ is at least $c_0+1$. Otherwise the CS where a node with at most $c_0$ neighbours is alone is a minimal cut and yields $m$ a utility of $|N(m)|-1$ or $|N(m)|$ in contradiction to inequality~\ref{ineq:util_ub_safe}.

We distinguish between multiple cases. 
Case one: $B_2\subseteq N$, i.e. contains only true neighbours of $m$.
Since $P^m$ has higher utility for the manipulator than all previous solutions, in all previous solutions there are at least $|B_2|+1$ neighbours of her not in her coalition. Hence $c_0 \geq |B_2|+1$. By the previous statement we get that $\delta(\overline{G}) \geq |B_2| + 2$. Since the manipulator might have removed one edge in $\overline{G}^m$ , we get that $\delta(\overline{G}^m) \geq |B_2| + 1$. Because of that, in $P^m$ every node in $B_2$ has at least $2$ neighbours outside $B_2$, which we concludes that $c_m \geq 2|B_2| \rightarrow c_0 > 2|B_2| \rightarrow \delta(\overline{G}) > 2|B_2| + 1 \rightarrow \delta(\overline{G}^m) \geq 2|B_2|+1$. Hence all nodes in $B_2$ has at least $|B_2|+2$ edges going out from $B_2$, so we get that $c_m > |B_2|*(|B_2|+2)$. By repeating this process we get that the cut is unbounded, in contradiction since the network is finite.

Case two: $|B_2\cap N| = |B_2| -1$, i.e. $B_2$ contains exactly one non-neighbour of $m$, denote her by $a$.
By the same logic from above we get that $c_0 \geq |B_2|$ (as in $P^m$ the manipulator only has $|B_2| -1$ neighbours. So $\delta(\overline{G}) \geq |B_2| + 1 \rightarrow \delta(\overline{G}^m) \geq |B_2|$. Since $a$ is not a neighbour of $m$, her degree in $\overline{G}^m$ has not changed and it is still at least $|B_2| +1$. In conclusion in $B_2$ we have $|B_2|-1$ nodes of degree at least $|B_2|$ and one with at least $|B_2| +1$. Hence $c_m \geq |B_2| +1 \rightarrow c_0 \geq |B_2| + 2$, and yet again by repeating the process we get that the cut is unbounded.

Case three: $B_2$ contains at least 2 non-neighbours of $m$.

This is the case we could not prove. Note that in any solution in Max-Util any agent is guaranteed to be in the coalition where she has more neighbours. Therefore if $B_2$ contains at least $2$ neighbours of $m$, then $B_1$ contains at least $3$. However, if the manipulator had only $5$ neighbours originally, he would have had at least $3$ neighbours before the manipulation. So he has at least $4$ neighbours in $B_1$.

To sum it up, if a weak-improvement did exist, it has to be that $B_1$ contains at least $4$ neighbours of $m$, $B_2$ contains at least $2$ neighbours and $2$ non-neighbours of $m$, and the minimum degree of the graph is at least $4$.